# Convergence Time Evaluation of Algorithms in MANETs

Annapurna P Patil          Narmada Sambaturu          Krittaya Chunhaviriyakul
Department of Computer Science and Engineering,
M.S. Ramaiah Institute of Technology, Bangalore-54, India.
annapurnap2@yahoo.com          narmada.sambaturu@gmail.com          krittaya_chun@hotmail.com

*Abstract*

Since the advent of wireless communication, the need for mobile ad hoc networks has been growing exponentially. This has opened up a Pandora's Box of algorithms for dealing with mobile ad hoc networks, or MANETs, as they are generally referred to.

Most attempts made at evaluating these algorithms so far have focused on parameters such as throughput, packet delivery ratio, overhead etc. An analysis of the convergence times of these algorithms is still an open issue. The work carried out fills this gap by evaluating the algorithms on the basis of convergence time.

Algorithms for MANETs can be classified into three categories: reactive, proactive, and hybrid protocols. In this project, we compare the convergence times of representative algorithms in each category, namely Ad hoc On-Demand Distance Vector (AODV)-reactive, Destination Sequence Distance Vector protocol (DSDV)-proactive, and Temporally Ordered Routing Algorithm (TORA)-hybrid.

The algorithm performances are compared by simulating them in ns2. Tcl is used to conduct the simulations, while perl is used to extract data from the simulation output and calculate convergence time. The design of the experiments carried on is documented using Unified modeling Language. Also, a user interface is created using perl, which enables the user to either run a desired simulation and measure convergence time, or measure the convergence time of a simulation that has been run earlier.

After extensive testing, it was found that the two algorithms AODV and DSDV are suited to opposite ends of the spectrum of possible scenarios. AODV was observed to outperform DSDV when the node density was low and pause time was high (sparsely populated very dynamic networks), while DSDV performed better when the node density was high and pause time was low (densely populated, relatively static networks). The implementation of TORA in ns2 was found to contain bugs, and hence analysis of TORA was not possible.

Future enhancements include rectifying the bugs in TORA and performing convergence time analysis of TORA as well. Also, a system could be developed which can switch between protocols in real time if it is found that another protocol is better suited to the current network environment.

## I. INTRODUCTION

Recently there has been tremendous growth in the number of laptops and mobile phones. With the increase in their number, their participation in people's basic needs to share information has also grown. In areas in which there is little or no communication infrastructure or the existing infrastructure is expensive or inconvenient to use, wireless mobile users will still be able to communicate through the formation of a Mobile Ad Hoc Network (MANET).

In a MANET, each node participates in an ad hoc routing protocol that allows it to discover multi-hop paths through network from itself to destination.

There are three types of routing protocols used in MANETs namely: proactive, reactive and hybrid routing protocols.

Each type of protocols performs differently under different network scenarios. One protocol might perform better than others in specific situation.

To compare performance of each type of protocol, a representative routing protocol is chosen from each type, namely: DSDV (proactive), AODV (reactive), and TORA (hybrid). These protocols are compared in terms of convergence time to uncover in which situations these types of algorithms have their strengths and weaknesses.

## II. RELATED WORK

Extensive work has been done on evaluating algorithms for MANETs. In [16], AODV and DSDV have been compared with average throughput, packet loss ratio, and routing overhead as the evaluation metrics, [17] has compared AODV and DSDV in terms of delay and drop rate, [18] compares AODV and DSDV in terms of throughput, packets received, delay and overload. Similarly, [19] compares AODV, DSDV and DSR in terms of throughput, delay, drop rate. This paper has also varied the mobility model used as Random Waypoint and Social Networking models.

## III. OVERVIEW OF THE DOCUMENT

Section 1 gives a general introduction to the project while section 2 explores related work. An overview of MANETs is given in section 4, along with a brief explanation of the chosen algorithms. Section 5 gives the system design and Implementation. Section 6 describes the testing of the system.

Conclusion and future enhancements is provided in Section 7 followed by references.

## IV. BACKGROUND

### A. MANETs

In MANETs, each mobile device operates as a host and also as a router [10], forwarding packets for other mobile device when they are not within direct wireless range of each other.

Each node participates in an ad hoc routing protocol that allows it to discover multi-hop paths through network from itself to a destination. MANET is an infrastructure less network [8] since the mobile nodes in the network dynamically establish routing among themselves to form their own network on the fly. Ad hoc networks are very different from wired networks since they have many constraints which are not present in wired networks e.g. limited bandwidth, limited battery and dynamic topology. Since the nodes are mobile, the network topology may change rapidly and unpredictably over time. The network is decentralized, and all network activity including discovering the topology and delivering messages must be executed by the nodes themselves, i.e., routing functionality is incorporated into the mobile nodes.

Since the topology in a MANET environment is highly dynamic, the path information the routing algorithm may have acquired can become invalid. The algorithm has to detect that the information it possesses is invalid, and find the new correct path to the destination. The time between a fault detection, and restoration of a new, valid path, is referred to as convergence time [6]. This is a very important metric in MANET scenarios, and reflects the ability of the algorithm to adapt to the changing network dynamics.
The algorithms must aim to minimize convergence time as much as possible.

There are three types of routing protocols used in MANETs: proactive, reactive and hybrid routing protocol. Proactive protocols are the ones which maintain routing information for all mobile nodes and keep this updating information periodically. An example of proactive protocols is DSDV (Destination Sequenced Distance Vector). Reactive protocols are the protocols which do not maintain routing information. Rather, they discover the path to destination on demand. This saves memory, battery power, and bandwidth. An example is AODV, or Ad hoc On demand Distance Vector. Hybrid protocols combine the advantages of both proactive and reactive protocols. An example is TORA, or Temporally Ordered Routing Algorithm.

### B. DSDV

DSDV is a table-driven routing scheme for ad hoc mobile networks based on the Bellman-Ford algorithm. The main contribution of the algorithm was to solve the Routing Loop problem which is present in Bellman-Ford algorithm. To do so, DSDV makes use of sequence numbers. Each entry in the routing table contains a sequence number; the sequence numbers are generally even if a link is present; else, an odd number is used. The number is generated by the destination, and the emitter needs to send out the next update with this number. Routing information is distributed between nodes by sending full dumps infrequently and smaller incremental updates more frequently.

### C. AODV

AODV is a reactive routing protocol, meaning that it establishes a route to a destination only on demand. When there is no need of a connection, the network remains silent. It is a distance-vector routing protocol. AODV avoids the count-to-infinity problem by making use of sequence numbers.

When a connection is needed, the network node that needs the connection broadcasts a request for finding a route to the destination .Other nodes forward the message, and record the id of the node that they heard it from, creating temporary routes back to the needy node. When a node receives such a message and already has a route to the destination, it sends a message backwards through a temporary route to the requesting node. The needy node then begins using the route that has the least number of hops to the destination. Unused entries in the routing tables are recycled after a time. When a link fails, a routing error is passed back to a transmitting node, and the process repeats.

### D. TORA

TORA is a hybrid protocol which combines the advantages of both proactive and reactive protocols. TORA does not use a shortest path solution. It builds and maintains a Directed Acyclic Graph rooted at a destination. No three nodes may have the same height. TORA makes use of height in order to prevent loops in routing. Information may flow from nodes with higher heights to nodes with lower heights, but not vice-versa [22].

## V. SYSTEM DESIGN AND IMPLEMENTATION

### A. Main design criteria

*Simulator chosen:* The network simulator ns2 [11] (version 2.33) was chosen, as this has become a de facto standard for networking research. It was necessary to use available implementations of algorithms rather than implement them freshly ourselves, as it is important for the acceptance of an evaluation that the implementation used for evaluation has been scrutinized and accepted as correct by the community. Else the evaluation results will not be accepted as doubt will exist about the correctness of the implementation of the algorithms.

*Algorithms chosen:* One representative algorithm from each type was chosen for comparison, as this would give a broad picture of which type of algorithm performs well in which environments. Further experiments can be built based on the results of this project, to compare convergence time performance of algorithms within the same category as well.

The specific algorithm chosen within each category was DSDV for proactive, AODV for reactive, and TORA for hybrid. These algorithms embody the principles of the type of algorithm they represent. Also, they have been implemented previously in ns2, and have been under community scrutiny for a long time. However it was found that ns2's implementation of TORA contains bugs [20. 21, 22], and forms loops. Hence only DSDV and AODV have been actually simulated. The hybrid algorithm has been left as a future enhancement.

*Mobility mode:.* The Random Waypoint Model was chosen as the mobility model for the simulations. According to this model, a node waits in its current position for a duration of time specified by pause time. At the expiration of this pause time, it chooses a destination randomly, and moves to it with a speed chosen from the uniform distribution [0, max_speed]. This process is repeated until the end of the simulation.

*Specific scenarios:* The parameters that define the MANET scenario are node density, and node mobility. In this project, the node density can be varied by varying the number of nodes, while the mobility can be varied by varying the pause time. That is, if a node pauses for a longer time at each waypoint, its overall mobility is less, while pausing for a very short time, say 1s, means its mobility is very high.

*B. Implementation*

- *Network scenario*

The simulations are conducted using the network simulator ns2 [11]. Random Waypoint mobility model is used. The physical layer simulates the behavior of IEEE 802.11 (as included with ns2). Each node has a radio range of 250 meter, and uses TwoRayGround as the radio propagation model.
All the scenarios are based on the following basic parameters:
cbr (constant bit rate) traffic
topology of size 500 m x 500 m
maximum speed of each node 20 m/s
simulation time 180s
transmission rate (packet rate) 10 m/s
The number of nodes is varied in the range [10,100] in steps of 10 (to represent 10 node densities). Pause time is varied in the range [0,180] in steps of 20 (to represent 10 pause times).

- *Convergence time measurement*

In [5], convergence time has been defined as the time between detection of an interface being down, and the time when the new routing information is available. [6] defines a route convergence period as the period that starts when a previously stable route to some destination becomes invalid and ends when the network has obtained a new stable route for. Similarly, we define convergence time as the time between a fault detection, and restoration of new, valid, path information.

[5] calculates convergence time in the IP backbone. The authors arrive at the value of convergence time by deploying entities called 'listeners', which listen to every link state PDU being sent by the is-is protocol. The time when the first 'adjacency down' packet is observed is taken as the time of detection of an interface being down. This failure event is said to end when the listener receives link state PDUs from both ends of the link.

We arrive at the convergence time by measuring the interval between the detection of route failure and successful arrival of a packet at the destination over the newly computed route. This includes not only the routing convergence time, but also the time taken for the packet to traverse the network from the source to the destination over the newly discovered path. Since this is a comparative analysis, and both the routing protocols use shortest distance with number of hops as the metric for distance calculation, both protocols will arrive at the same new route, and the time taken to reach the destination over this new route will be the same (since all physical characteristics are the same). Hence this extra time measured does not affect the comparative analysis.

In any case, the time taken for a packet to travel from the source to the destination is negligible when compared to the time taken for the algorithm to discover the new route, either through route request - route reply sequences as in reactive protocols, or by waiting for an update that contains new route information as in proactive protocols. Also, this automatically verifies that the new path calculated is correct.

The cycle of invalidation of old path and discovery of a new path might occur many times, and for many source-destination pairs over the course of the simulation. Hence the average value of these times is taken as the convergence time of that algorithm for that scenario.

This procedure has been carried out in perl.

VI. TESTING

In order to be able to cover most if not all the types of scenarios the algorithms might face, we varied both the node density (number of nodes) and the node mobility (pause time). The node density (number of nodes) was varied in the range [10,100] in steps of 10 (10 different node densities).

The upper limit of this range was chosen to be 180 because the simulation time is 180s in all the cases. Thus a pause time of 180 implies that the nodes pause in their initial positions for 180 seconds – the entire duration of the simulation. Hence this represents the case where nodes are completely static. Similarly, pause time 0 represents very high mobility where the nodes are in constant motion. Thus we tested each algorithm over 10 node densities x 10 pause times = 100 scenarios.

Also, each scenario was generated 3 times with the same parameters but with different seeds to the random number generators. This gave us a different set of connections and a different mobility signature each time. The convergence time was measured for each of these three scenarios, and the average of these was used in the final analysis. Thus every single point in the graphs shown below is the result of 3 simulations. At the conclusion of the project, a total of 600 simulations ((100 scenarios x 3 runs of each scenario) x 2 algorithms) had been run. All the graphs shown below plot the value of convergence time against pause time. In each graph, the node density is fixed.

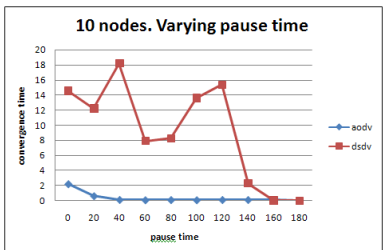
Figure 1. 10 nodes, varying pause time

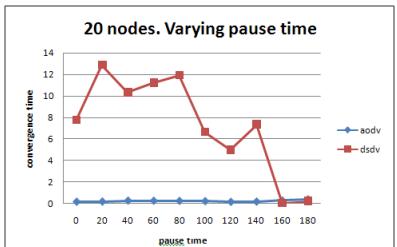
Figure 2. 20 nodes, varying pause time

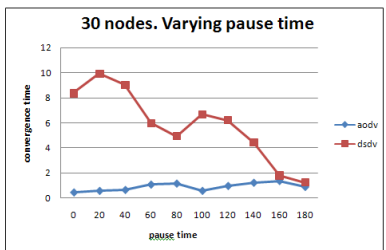
Figure 3. 30 nodes, varying pause time

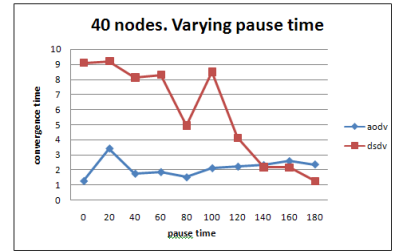
Figure 4. 40 nodes, varying pause time

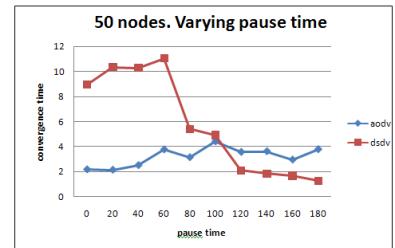
Figure 5. 50 nodes, varying pause time

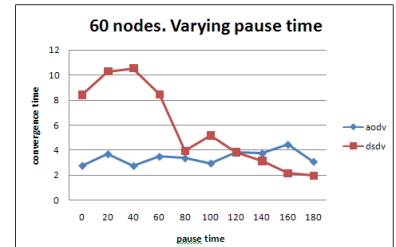
Figure 6. 60 nodes, varying pause time

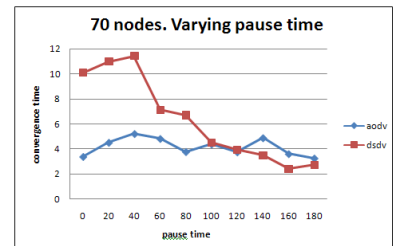
Figure 7. 70 nodes, varying pause time

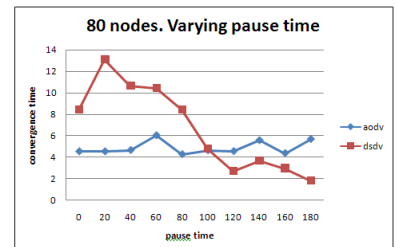
Figure 8. 80 nodes, varying pause time

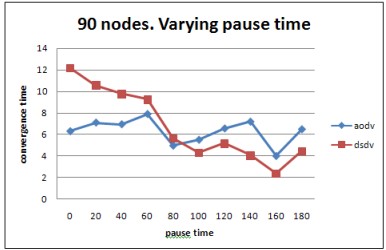

Figure 9. 90 nodes, varying pause time

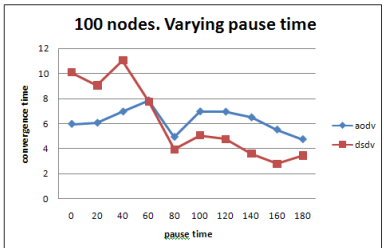

Figure 10. 100 nodes, varying pause time

Figures 1, 2 and 3 give the graphs of convergence times when the number of nodes were fixed to 10, 20 and 30, respectively (low node density). AODV was found to converge in less than 1 second most of the time, while DSDVs convergence characteristics varied with changes in the pause time. However, DSDV's performance matched that of AODV when pause time was 160s, and 180s.

Figures 4, 5, 6 and 7 describe the cases when there were 40, 50, 60 and 70 nodes in the scenario (intermediate node densities). DSDV was observed to converge faster than AODV at pause times of 140s, 160s, and 180s (when nodes are almost immobile). When the pause time was low (high mobility), AODV continued to outperform DSDV.

In figures 8, 9 and 10, the number of nodes in the scenario is 80, 90 and 100, respectively. DSDV was observed to converge faster than AODV in all cases except in extremely high mobility (at 0s, 20s, 40s pause times). AODV's convergence time degraded to its worst, standing at 8s for 100 nodes and pause time 60s.

## VII. CONCLUSION AND FUTURE ENHACEMENTS

### A. Conclusion

The aim of this project was to compare representative algorithms - DSDV (proactive), AODV (reactive) and TORA (hybrid) in terms of convergence time, and uncover in which situations these types of algorithms have their strengths and weaknesses.

Ns2 [11] was chosen to simulate these algorithms, and a perl script was written to measure convergence time by parsing the output of the simulations. The implementation of TORA in ns2 was found to contain bugs which caused it to form long living loops [20, 21, 22]. Hence an analysis of TORA was not possible. 600 simulations of the algorithms DSDV and AODV were conducted, over a wide range of network scenarios.

When the results of the simulations were analyzed, it was found that AODV was able to converge in less than 1s for scenarios where there was low node density (of the order of 30 nodes or lower) and high to very high mobility (of the order of 60 pause time or lower). In these scenarios, DSDV was found to perform very poorly, taking upto 18s to converge.

However, with increase in node density, the convergence time of AODV was found to increase steadily, while that of DSDV was found to decrease. For the case when node density was high (of the order of 80 nodes or higher) and mobility was low (of the order of 100s pause time or higher), DSDV was found to converge faster than AODV, with AODV taking as much as twice as long to converge.

For intermediate node densities (40 to 70 nodes) and intermediate mobilities (60s to 80s pause time), the convergence time performances of the two algorithms were found to be comparable (of the order of 4s).

Thus AODV and DSDV were together found to cover the two ends of the spectrum of possible network scenarios, with AODV providing fast convergence with low node densities and high mobilities, and DSDV performing well with high node densities and low mobilities.

### B. Future Enhancements

The project developed was a primitive attempt and can be further improved to include the following features:

- To correct the errors present in TORA routing protocol and perform convergence time analysis.
- Analysis of various other hybrid routing protocols could be experimented.
- The switching from one routing protocol to another routing protocol can be made possible when the current network scenario can perform better with another protocol. Hence in real time application, this switching mechanism can provide a better performance routing when the real time information is necessary.
- Different routing protocols from each type of MANET routing algorithms can be determined and compared to find the performance.

ACKNOWLEDGMENT

We wish to acknowledge the efforts of our Professor and Head of the Department at CSE Dr. V. K. Aanthashayana for his guidance which helped us work hard towards producing an innovative work.

AUTHORS PROFILE

The authors are Faculty and Students at M S Ramaiah Institute of Technology, Bangalore working in the area of Bio-inspired computing in the RD labs, Department of Computer Science.